# Tungsten based Anisotropic Metamaterial as an Ultra-broadband Absorber

Yinyue Lin,[1,4] Yanxia Cui,[1,*] Fei Ding,[2] Kin Hung Fung,[3] Ting Ji,[1] Dongdong Li,[4] and Yuying Hao[1,*]

[1]College of Physics and Optoelectronics, Taiyuan University of Technology, Taiyuan 030024, China

[2]Centre for Nano Optics, University of Southern Denmark, Campusvej 55, DK-5230 Odense M, Denmark

[3]Department of Applied Physics, The Hong Kong Polytechnic University, Hong Kong

[4]Shanghai Advanced Research Institute, Chinese Academy of Sciences, 99 Haike Road, Zhangjiang Hi-Tech Park, Pudong, Shanghai 201210, China

[*]yanxiacui@gmail.com, hyy_123991@sina.com

## ABSTRACT

The trapped rainbow effect has been mostly found on tapered anisotropic metamaterials (MMs) made of low loss noble metals, such as gold, silver, etc. In this work, we demonstrate that an anisotropic MM waveguide made of high loss metal tungsten can also support the trapped rainbow effect similar to the noble metal based structure. We show theoretically that an array of tungsten/germanium anisotropic nano-cones placed on top of a reflective substrate can absorb light at the wavelength range from 0.3 $\mu$m to 9 $\mu$m with an average absorption efficiency approaching 98%. It is found that the excitation of multiple orders of slow-light resonant modes is responsible for the efficient absorption at wavelengths longer than 2 $\mu$m, and the anti-reflection effect of tapered lossy material gives rise to the near perfect absorption at shorter wavelengths. The absorption spectrum suffers a small dip at around 4.2 $\mu$m where the first order and second order slow-light modes get overlapped, but we can get rid of this dip if the absorption band edge at long wavelength range is reduced down to 5 $\mu$m. The parametrical study reflects that the absorption bandwidth is mainly determined by the filling ratio of tungsten as well as the bottom diameter of the nano-cones and the interaction between neighboring nano-cones is quite weak. Our proposal has some potential applications in the areas of solar energy harvesting and thermal emitters.

In the past decade, metamaterials (MMs) [1, 2] which are composed of artificial functional subwavelength inclusions, have gained extensive attention due to their fascinating electromagnetic properties. Many exotic phenomena such as super-lenses [3], cloaking [4], anomalous light bending [5, 6], near field concentration [7], asymmetric light transmission [8] and so on, have been realized with MMs. Although metal losses affect some performances [9], MMs have shown great advantages at exerting perfect electromagnetic absorption [10-14]. Ever since the pioneer proposal of absorbers based on tapered anisotropic MMs comprising of alternating metal and dielectric plates was firstly put forward in 2012 [15], it has never been easier to realize electromagnetic absorption with near-unity intensity over a broadband frequency range, facilitating the potential applications of absorbers in solar energy harvesting, thermal emitters, infrared bolometers, as well as military defense systems. Such a tapered anisotropic MM slab, showing the hyperbolic dispersion diagram, enables the collective excitation of slow-light waveguide modes at a broadband wavelength range, also known as the trapped rainbow effect [16-19]. Soon after it being firstly proposed at the middle-infrared region in theory [15], the exact concept was experimentally demonstrated at microwave [20, 21], terahertz [22] as well as optical frequencies [23-26].

It is noticed at optical frequencies, most of the reported broadband absorbers based on anisotropic MMs have employed the noble metal, such as gold, silver, etc., as one ingredient of the alternating layers [23-30]. In this work, we propose an unprecedentedly ultra-broadband electromagnetic absorber which can cover the ultraviolet, visible, near-infrared and even part of the mid-infrared wavelength regions. The approach is to employ tungsten, a high loss metal, replacing noble metal to form the tapered anisotropic MMs. Materials such as high loss metals, low bandgap semiconductors, etc., are compelling choices for designing absorbers, which have already aroused increasing attentions in the area of solar-thermal conversion [31-36]. The adjustability of designing absorbers based on stratified films is low [32, 33, 37], therefore many efforts have been made to structure the surfaces of high loss materials for designing absorbers with better spectral performances [31, 38, 39]. For example, the pyramid shaped tungsten (W) surface has been frequently investigated [40-43] and a germanium (Ge) periodic nanopyramid array has also been explored [36]. However, without the help of MMs, the absorption bands of pyramid tungsten with efficiency higher than 90% are limited to the wavelength range shorter than 2 $\mu$m [40-43] and that of pyramid germanium is even narrower [36].

Instead of using the bulk tungsten or germanium to construct the pyramid surface, we propose to corrugate a multilayer slab comprising of alternating tungsten and germanium thin films into a two-dimensional (2D) array of nano-cones, which demonstrates the near unity absorption over the wavelength band between 300 nm and ~10 $\mu$m, approaching the recently reported results by three-dimensional (3D) self-assembly of silver nanoparticles [13]. Our study also shows that the trapped rainbow effect originally proposed in plasmonic structures made of noble metals [16-19] can be observed in high loss metal based structure. The delicate design enables the existence of slow-light waveguide modes of not only the fundamental order but also other higher orders. The band of the absorber with only the fundamental order slow-light mode is already very broad [15]; here, through carefully tuning the tapered geometry, the absorption bands produced by different orders of slow-light waveguide modes are linked to one another, displaying an unprecedentedly ultra-broadband performance. Such outstanding performance can hardly be realized using either bulk tungsten or germanium. Another merit of the present design lies on the performance at the wavelength band between 300 nm and 2 $\mu$m, where either pyramid tungsten or germanium already shows near unity absorption [36, 40]. In addition, tungsten have a high melting point, so it can withstand high temperature which is requisite in thermo-photovoltaic and other high temperature applications [44].

Figure 1 shows the schematic diagram of our proposed 3D anisotropic MM absorber comprising of a 2D periodic array of nano-cones which is formed by alternating tungsten and germanium thin films. In total, there are $N$ pairs of tungsten and germanium films with thicknesses of $t_1$ and $t_2$, respectively. The top and bottom diameters of the nano-cones are $W_1$ and $W_2$, respectively, and the height of cones is $H$. The periodicity of the nano-cone array along both $x$ and $y$ directions are $P$. A gold film with a thickness of $t$ = 200 nm is placed under the nano-cones to block any light transmission. Full-wave electromagnetic simulations are performed with a commercial software of Lumerical based on the finite-difference time-domain (FDTD) method. In the simulations, a polarized plane wave with the magnetic field polarized along $y$ axis (denoted as TM polarization) impinges normally on the structure unless stated otherwise. The dispersive permittivities of tungsten, germanium and gold are taken from Ref. [45]. The absorptivity is calculated by $A(\lambda) = 1 - R(\lambda)$, where $R(\lambda)$ represents the reflectivity.

For a proposed W/Ge nano-cone array on top of a reflective substrate with geometrical parameters of $t_1 = t_2$ = 35 nm, $W_1$ = 200 nm, $W_2$ = 1000 nm, $H$ = 2100 nm, and $P$ = 1000 nm, the absorption at normal incidence is efficient at a quite broad wavelength band as shown by the curve w/o symbols in

Fig. 2(a). The absorption efficiency at the wavelength range between 300 nm and 9 μm is almost perfect, except baring a small dip at 4.2 μm with amplitude of ~80%, corresponding to an integrated absorptivity of 98%. For comparison, both bulk W and bulk Ge nano-cone arrays on top of a reflective substrate with exactly the same geometrical parameters of $W_1$, $W_2$, $H$, and $P$ are also investigated with their absorption spectra as shown in Fig. 2(b) and 2(c), respectively. Compared with our proposed absorber, either bulk W or bulk Ge based structure displays quite narrower absorption band. Obviously, the bulk W nano-cone array shows the poorest overall absorption performance whereas that of the bulk Ge nano-cone array is a bit better. There is no doubt that the absorption of our proposed structure shows great superiority in terms of both bandwidth and absorbance. Another control sample with only the free-standing W/Ge nano-cone array is also investigated with its absorption spectrum displayed in Fig. 2(d). It is clearly observed that the bottom reflective substrate favors broadening the absorption band especially at the long wavelength range (7-12 μm). In practical applications such as solar energy harvesting, it is desirable that the absorption performance is insensitive to the incident angle along with the polarization angle of the incidence. Here, for the proposed W/Ge nano-cone array on top of a reflective substrate, we calculated its absorption at oblique incidence for both TM and TE (with the electric field of the incident light polarized along *y* direction) as shown in Fig. S1. It indicates that the demonstrated broadband absorption effect is quite insensitive to the incident angle. The absorption efficiency still remains above 70% over the investigated wavelength band even when then incident angle reaches 60 degrees for both polarized incident cases.

Next, we apply the effective medium theory to understand the physical mechanism of the ultra-broadband absorption of the proposed absorber. An alternate-layered structure made of material 1 and material 2 can be regarded as a homogeneous medium with effective anisotropic relative permittivities of $\varepsilon_\parallel$ and $\varepsilon_\perp$ when the thickness of each layer is much smaller than the wavelength according to the following equation,

$$\varepsilon_\parallel = \varepsilon_x = \varepsilon_y = f \cdot \varepsilon_1 + (1-f) \cdot \varepsilon_2$$
$$\varepsilon_\perp = \varepsilon_z = \frac{\varepsilon_1 \cdot \varepsilon_2}{f \cdot \varepsilon_1 + (1-f) \cdot \varepsilon_2} \tag{1},$$

where $\varepsilon_1$ and $\varepsilon_2$ represent the relative permittivities of material 1 and material 2, and *f* is the filling ratio of material 1. The wavelength dependent relative permittivities of the effective anisotropic medium constructed by alternating W and Ge films with a filling ratio of $f = 0.5$ (as in the proposed absorber, both W and Ge films are 35 nm thick) are shown in Fig. S2. If there is no surface pattern carved on such an anisotropic medium slab with a thickness of $H$ placed on top of a reflective substrate, energy harvesting cannot be realized as shown in Fig. S3(a). However, if the anisotropic effective layer patterned one-dimensionally into nano-sawteeth or two-dimensionally into, e.g., nano-cones, the reflection can fully vanish over a broad wavelength range. Based on the obtained anisotropic permittivities, we calculated the absorption spectrum of the homogeneous anisotropic nano-cones on top of a reflective substrate as shown by the curve w/ symbols in Fig. 2(a). It is found that the performance of the effective model is in good agreement with that of the real absorber, indicating that that the effective medium theory used here is valid. Such good agreement with the effective medium theory also suggests that the properties could be robust against imperfection in some fabrication errors in making the multi-layers. Fig. S3(b) displays the absorption spectrum of an array of anisotropic sawteeth on top of the reflective substrate under TM polarization; the similar ultrabroadband absorption effect is produced.

For the present high loss metal based absorber, we need to figure out whether it can excite slow-light modes. The nano-cones based absorber have a circular-shaped cross section similar to a circular waveguide core. Here, a simplified model, a 2D waveguide with a W/Ge anisotropic MM slab (as the core) embedded in air claddings, is investigated. Such a simplification is viable as we see the absorption spectra of the nano-cones [Fig. 2(a)] and the nano sawteeths [Fig. S3(b)] under TM polarization are similar and there is no high order oscillation of the magnetic field along $y$ direction according to the field maps at the cross section planes of the nano-cones [Fig. S4]. For the 2D W/Ge anisotropic waveguide, we analytically solved the dispersion relationship between the incident photon frequency ($\omega_c = \omega/c$) and the propagating constant ($\beta$) with a fixed core width of $W$ according to the following equation

$$\exp[iq_2 W] + \frac{\kappa_1 - i\frac{q_2}{\varepsilon_\perp}}{\kappa_1 + i\frac{q_2}{\varepsilon_\perp}} = 0 \qquad (2),$$

where $\kappa_1 = (\beta^2 - \omega_c^2)^{1/2}$ and $q_2 = \left(\varepsilon_\perp \omega_c^2 - (\varepsilon_\perp/\varepsilon_\parallel)\beta^2\right)^{1/2}$. Dispersion curves of fundamental and high order waveguide modes of the W/Ge anisotropic waveguide (with the relative permittivities as shown in Fig. S2) at different core widths are displayed in Fig. 3(a)-3(c). One can clearly see that for each dispersion curve, the mode cuts off at a certain frequency and there is a degeneracy point where the group velocity ($v_g = d\omega_c/d\beta$) approaches zero. In other words, the present high loss metal based anisotropic waveguide can excite the slow-light modes. Wavelength points with $v_g = 0$ for the waveguide modes of different orders are extracted when the core width $W$ is tuned, as plotted in Fig. 3(d). It indicates that the wavelength of slow-light modes gets longer with the increase of $W$. The tapered waveguide with $W$ between 200 nm ($W_1$) and 1000 nm ($W_2$) can support fundamental slow-light modes [squares, Fig. 3(d)] in the spectrum ranging from 2.7 $\mu$m to 11.7 $\mu$m. When $W$ is larger than 400 nm, a higher order slow-light mode [circles, Fig. 3(d)] which can cover the wavelength range from 2 $\mu$m to 4 $\mu$m can be excited. Another higher order modes can be excited when $W$ is larger than 800 nm and the corresponding wavelength range is between 2.1 $\mu$m and 2.5 $\mu$m [triangles, Fig. 3(d)]. All these modes can be found in the field distribution maps.

Fig. 3(e)-3(j) plot the normalized magnetic field distributions at the plane of $y = 0$ at the incident wavelength of $\lambda = 9$ $\mu$m, 5 $\mu$m, 4 $\mu$m, 3 $\mu$m, 2.5 $\mu$m, and 2.1 $\mu$m, respectively. Obviously, we can see that at long wavelength of $\lambda = 9$ $\mu$m, there is a fundamental order slow-light mode in the form of a bright spot existing at the bottom of nano-cones. When the wavelength becomes shorter, the bright spot goes up to the middle part of nano-cones (at $\lambda = 5$ $\mu$m) and finally arrives at the cone's head (at $\lambda = 3$ $\mu$m). This is exactly the trapped rainbow effect, as what reported in the noble metal based broadband absorber [15]. The observed phenomenon of the field distributions are in good agreement with the dispersion diagram of the fundamental order waveguide mode. If only the fundamental order of slow-light waveguide modes are excited, the bandwidth of absorption could not be as ultra-broad as observed in Fig. 2(a). This is because at the top and bottom faces of the waveguide, the bright spot cannot get fully expanded and the distortion of the mode profile reduces the absorption. For example, as one see in Fig. 3(h), when the fundamental order slow-light mode approaches the top face of the waveguide, only the bottom half of the bright spot is witnessed, which eventually weakens the absorption of light. Similar effect happens at the bottom of the nano-cones (not shown), that is why after removing the bottom reflective substrate, the absorption at long wavelength band (7-12 $\mu$m) is severely degraded [Fig. 2(d)]. By using the reflective substrate, light at long wavelengths can travel back into the anisotropic waveguide, instead of transmitting through, and then get fully absorbed, thereby the band edge with perfect absorption can be extended to 9 $\mu$m. For the degraded absorption

at the top face of the nano-cone, exciting high order slow-light waveguide modes can be a solution.

Here, we see from Fig. 3(d) that high order slow-light modes can be excited in the waveguides with large $W$ and the corresponding wavelengths are shorter than the fundamental one. In addition, there is an overlap between the wavelength bands of different orders of modes. At $\lambda = 4$ $\mu$m, we see from Fig. 3(g) that, beside of the fundamental order of slow-light mode at the top part of nano-cones, there is also three bright spots emerging at the cone's bottom. Similar to the fundamental order, when the wavelength gets shorter, these three bright spots gradually move up, but they can only reach the cone's middle-upper part [Fig. 3(j)] with the cone diameter of around 400 nm, relating to the smallest core width that allows the excitation of this order of waveguide mode as shown in Fig. 3(b). By counting the number of the bright spots, we know that this mode belongs to the third order slow-light waveguide mode. It is noted that even order modes in anti-symmetrical form are dark modes, which cannot be excited under normal incidence [46]. We see in Fig. 3(i) and 3(j) that, at $\lambda = 2.5$ $\mu$m and 2.1 $\mu$m, the fifth order slow-light modes (which have five bright spots) are excited at the bottom of the cone. According to Fig. 3(c), this mode can only stay at the cone's bottom with waveguide core width greater than 800 nm. By merging different orders of slow-light waveguide modes together, a continuous absorption band ranging from 2.1 $\mu$m to 12 $\mu$m has been produced though there is still a tiny dip at $\lambda \sim 4.2$ $\mu$m due to the imperfect overlap of the fundamental and second order slow-light modes. We can completely get rid of this dip if the band width is sacrificed a bit. As shown in Fig. S5(a), for an absorber with $W_1 = 100$ nm, $W_2 = 600$ nm, and other geometrical parameters not changed, we can achieve perfect absorption when $\lambda$ is shorter than 5.2 $\mu$m.

The above analysis has ascribed the high efficiency absorption at wavelength longer than 2 $\mu$m to the trapped rainbow effect. At short wavelengths, since there is no slow-light waveguide modes excited, the mechanism of the observed high efficiency absorption must be examined otherwise. The top panel of Fig. 4 plots the normalized magnetic field distributions at the plane of $y = 0$ at the incident wavelength of $\lambda = 1.5$ $\mu$m, 1 $\mu$m, and 0.5 $\mu$m, respectively. It is seen that when the wavelength is shorter than 2 $\mu$m, the field distributions are evidently distinct from those at long wavelengths. Instead of staying within the W/Ge anisotropic waveguide, the interference fringes of light mainly take place outside of the W/Ge waveguide, i.e., in the air gap between neighboring nano-cones. Beside of the interference effect between incident and reflected light along $z$ axis, the side-boundaries of neighboring nano-cones also impact the profiles of fringes along the horizontal plane. It is seen in Fig. 4(a) and Fig.

4(b) that when the air gap is small (near the bottom of the nano-cones), the interaction between neighboring nano-cones is strong, resulting in strong concentration of light in the gap as well as some apparent penetration of energy into the nano-cones. For a large air gap that locates close to the top of the nano-cones, the interaction between neighboring nano-cones becomes weak, so that the interference fringes tend to be strong at the side-boundary of the cones but weak at the center of the air gap. As the wavelength becomes shorter, e.g., at $\lambda = 0.5$ $\mu$m, the number of the interference fringes along $z$ axis increases and the profiles of fringes along $x$ axis become more complicated.

From the afore-presented study [Fig. 2(b)], we know that the bulk W nano-cone array also shows efficient absorption at the short wavelength range, so we plot its field distributions at the corresponding wavelengths in the bottom panel of Fig. 4 as well. By comparing the top and bottom panels of Fig. 4, we find that the field distributions of W/Ge anisotropic and bulk W nano-cones are more or less the same, implying that their mechanisms of absorption should be the same. At short wavelengths, the real part of the relative permittivities of bulk W is positive, therefore the W nano-cone acts like a lossy dielectric which can naturally anti-reflect light due to its tapered profile. However, at long wavelengths, the real part of the relative permittivities of bulk W turns to be negative, which is several times greater than its imaginary part, leading to the W nano-cones show the metal property, thereby severely reflecting light. Here, the W/Ge anisotropic nano-cones confront similar situations. As shown by the enlarged plot of Fig. S2(a), with the shortening of the wavelength, the real part of the parallel component of the relative permittivity [Re($\varepsilon_\parallel$)] changes from negative to positive at $\lambda = \sim 1.6$ $\mu$m. At short wavelengths, although the positive Re($\varepsilon_\parallel$) of the W/Ge anisotropic slab leads to the failure to get any slow-light mode solutions from Eq. (2), the W/Ge nano-cones can still perform like high loss dielectric tapers, similar to bulk W cones, therefore light can be efficiently harvested due to the anti-reflection effect. Until now, the mechanisms of absorption at the ultra-broadband wavelength range have been clarified. It is the combination of the trapped rainbow effect of anisotropic MM waveguides and the anti-reflection effect of tapered lossy dielectric creates the proposed ultra-broadband and high efficiency absorber.

The absorption band edge at the long wavelength range is mainly determined by the resonance of the fundamental order slow-light waveguide mode. As derived from Eq. (2), the resonant wavelength of the fundamental order slow-light mode is approximately proportional to the square root of $\varepsilon_\perp$ and the waveguide width [15]. Thus, the absorption bandwidth can be easily tuned by changing the material proportion along with the nano-cone diameter. Fig. 5(a) shows the absorption spectra when the ratio of $t_1$ (the thickness of W plate) over $t_2$ (the thickness of Ge plate) is tuned. It is seen that the more Ge inclusion, the narrower the absorption band becomes; the more W inclusion, the broader the absorption band becomes. This is because with the increase of Tungsten in the composite film, the perpendicular permittivity of the anisotropic film increases, thereby the resonant wavelength of the fundamental slow-light mode is longer. However, if the bandwidth is over large (e.g., with $t_1$: $t_2$ = 4:1), more than one obvious absorption dips are produced, leading to inefficient harvesting of the broadband incident light. Such absorption dips can also be observed when the height of the nano-cones (with the diameters of the nano-cones fixed) decreases as shown in Fig. S5(b). As reported in Ref. [47], the adiabatic condition is necessary for the gap surface plasmon to achieve broadband nano-focusing by cancelling completely the mode reflection. Here, similarly, when the MM waveguide along the mode propagation direction changes dramatically due to either index or geometry change, significant reflection would take place. For the present W/Ge nano-cones, a profile of convex taper, satisfying the adiabatic condition, must be better than a linearly taper in terms of the broadband absorption effect. With the decrease in the bottom diameter of the nano-cones, the absorption band width becomes narrower as shown in Fig. 5(b), while the decrease in the top diameter of the nano-cones only influences the depth of the absorption dip around 4.2 $\mu$m as shown in Fig. 5(c). In addition, it is found in Fig. 5(c) that, as $W_1$ decreases, the absorption at the short wavelength range also degrades due to a relatively smaller filling ratio of the nano-cone per period. We also show the absorption spectra of the present absorber when the periodicity of the nano-cones is tuned in Fig. 5(d). It is seen that the periodicity does not affect the centers as well as the shapes of the absorption bands produced by different types of resonances. When $P$ increases from 1000 nm to 1300 nm, the absorption efficiencies over the whole investigated wavelength range get lower due to the reduced filling ratio of the cone, reflecting that the interaction between neighboring nano-cones is quite weak. This is different from that reported in Ref. [24], in which the center of the absorption band shifts due to the interaction between neighboring MM waveguides at the bottom side. All these parametrical studies display consistent conclusion that the underground mechanism of the demonstrated absorption is produced by the slow-light excitations at the long wavelength range and the antireflection effect at the short wavelength range due to individual W/Ge anisotropic nano-cone. The collective effect of an array of nano-cones only influences the intensity of the absorption.

In conclusion, we have designed an electromagnetic absorber that is broadband, wide-angle and insensitive to the polarization of the incident light based on an array of W/Ge nano-cones on top of a reflective substrate. It can absorb the whole incidence from 0.3 $\mu$m to 9 $\mu$m with an average absorption efficiency approaches 98%. As the high order slow-light modes join in harvesting the incident light together with the fundamental order mode, the absorption at the wavelength range longer than 2 $\mu$m can be broad and efficient. With the utilization of high loss metal tungsten, instead of noble metal, the absorption at the wavelength range shorter than 2 $\mu$m is very high due to the anti-reflection effect. The average absorption can be close to unity if the absorption band edge at the long wavelength range is sacrificed a bit to 5.2 $\mu$m by decreasing the diameters of the nano-cones. The designed electromagnetic absorber could be promising in applications such as optical detectors, solar energy harvesting, and thermal emitters.

## ACKNOWLEDGEMENT

National Natural Science Foundation of China (NSFC) (61475109, 61274056, and 61474128); Key Research and Development (International Cooperation) Program of Shanxi (201603D421042); Hong Kong RGC (AoE/P-02/12); Young Talents Program of Shanxi Province; Young Sanjin Scholars Program of Shanxi Province; the Outstanding Youth Funding at Taiyuan University of Technology.

**Figure Captions**

Figure 1 Schematic diagram of the proposed ultrabroadband absorber comprising of anisotropic MM nanocones on top of a reflective substrate. The top and bottom diameters of the nano-cones are $W_1$ and $W_2$, respectively, and their height is $H$. The nano-cones are periodically arrayed in the $x$-$y$ plane with periodicity of $P$. The MM is made of alternating tungsten and germanium thin films with thicknesses of $t_1$ and $t_2$, respectively. The reflective substrate is made of gold which is sufficiently thick to block light.

Figure 2 Absorption spectra of different structures. (a) W/Ge anisotropic nano-cone array on top of the reflective substrate. Solid line: the real structure. Symbolized line: the effective-medium structure. (b-c) Bulk W or Ge nano-cone array on top of the reflective substrate. (d) W/Ge anisotropic nano-cone array without the substrate. The diagram of the structural unit is also displayed in the inset. $t_1 = t_2 = 35$ nm, $W_1 = 200$ nm, $W_2 = 1000$ nm, $H = 2100$ nm, and $P = 1000$ nm.

Figure 3 (a-c) 1st, 3rd, and 5th order dispersion curves of the 2D W/Ge anisotropic waveguide when the width of the waveguide core ($W$) is tuned. (d) The wavelengths at which the slow-ight is excited (i.e., the group velocity approaches zero, $v_g = 0$) with tuned core width $W$ for the different orders of waveguide modes. (e-j) Distributions of the normalized magnetic field in the $x$-$z$ plane at $y = 0$ at the incident wavelengths of 9, 5, 4, 3, 2.5, and 2.1 $\mu$m.

Figure 4 Distributions of the normalized magnetic field in the $x$-$z$ plane at $y = 0$ for tungsten/germanium anisotropic nano-cone (a-c) and tungsten nano-cone (d-f) array on top of the reflective substrate at $\lambda = 1.5$, 1, and 0.5 $\mu$m, respectively.

Figure 5 Absorption spectra of the tungsten/germanium anisotropic nano-cone array on top of the reflective substrate when the ratio of $t_1$ over $t_2$ (a), the bottom diameter of nano-cones ($W_2$) (b), the nanocone height ($H$) (c), and the nano-cone periodicity ($P$) (d) are tuned.

Figure S1 Angular dependent absorption spectra of the W/Ge anisotropic nano-cone array on top of the reflective substrate at TM (a) and TE (b) polarizations.

Figure S2 Complex relative permittivities of the W/Ge alternating film ($t_1$:$t_2$ = 1:1). (a) The parallel component. (b) The perpendicular component. The inset of (a) shows the enlarged plot of the real part of the parallel component at wavelengths shorter than 3 $\mu$m.

Figure S3 Absorption spectra of W/Ge anisotropic film (a) and sawteeth (b) on top of the reflective

substrate. The diagram of the structural unit is also displayed in the inset.

Figure S4 Distributions of the normalized magnetic field at the *x-y* cross sections when $\lambda = 2.5$ μm. (a) $z = 1983$ nm, (b) $z = 1068$ nm, and (c) $z = 93$ nm, corresponding to the positions when the field amplitude of the first, third, and fifth order of slow-light modes is locally maximal as indicated in Fig. 3(i).

Figure S5 Absorption spectra of W/Ge anisotropic nano-cone array on top of the reflective substrate when $W_1$ and $W_2$ are equal to 100 nm and 600 nm, respectively (a), and $H = 900$ nm (b). Unless mentioned, other geometrical parameters are the default. $t_1 = t_2 = 35$ nm, $W_1 = 200$ nm, $W_2 = 1000$ nm, $H = 2100$ nm, and $P = 1000$ nm.

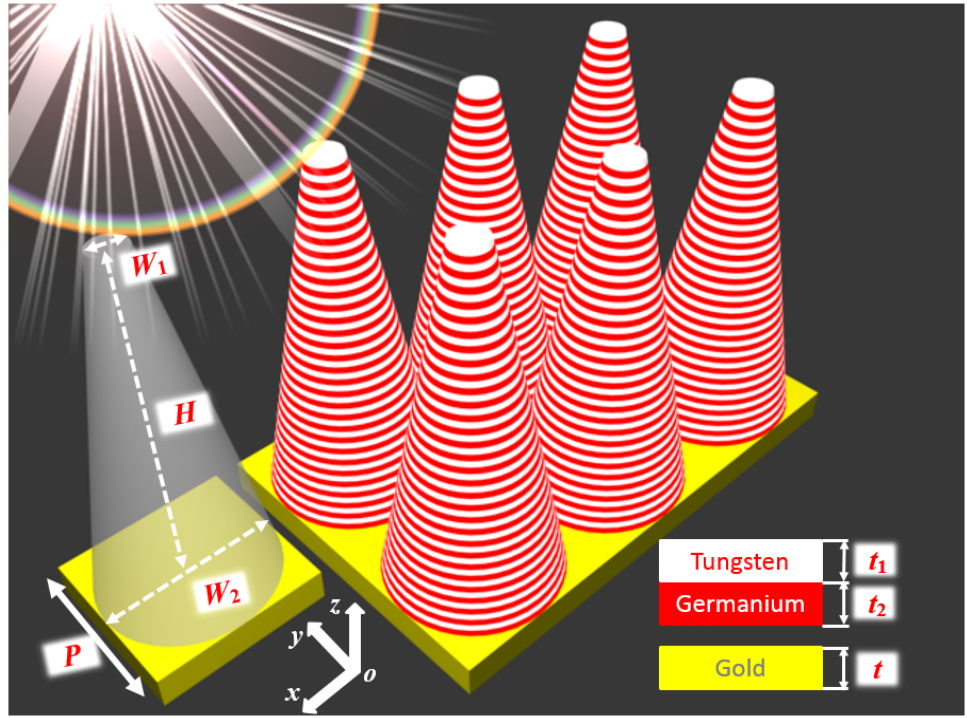

Fig. 1 Lin, Cui, et al.

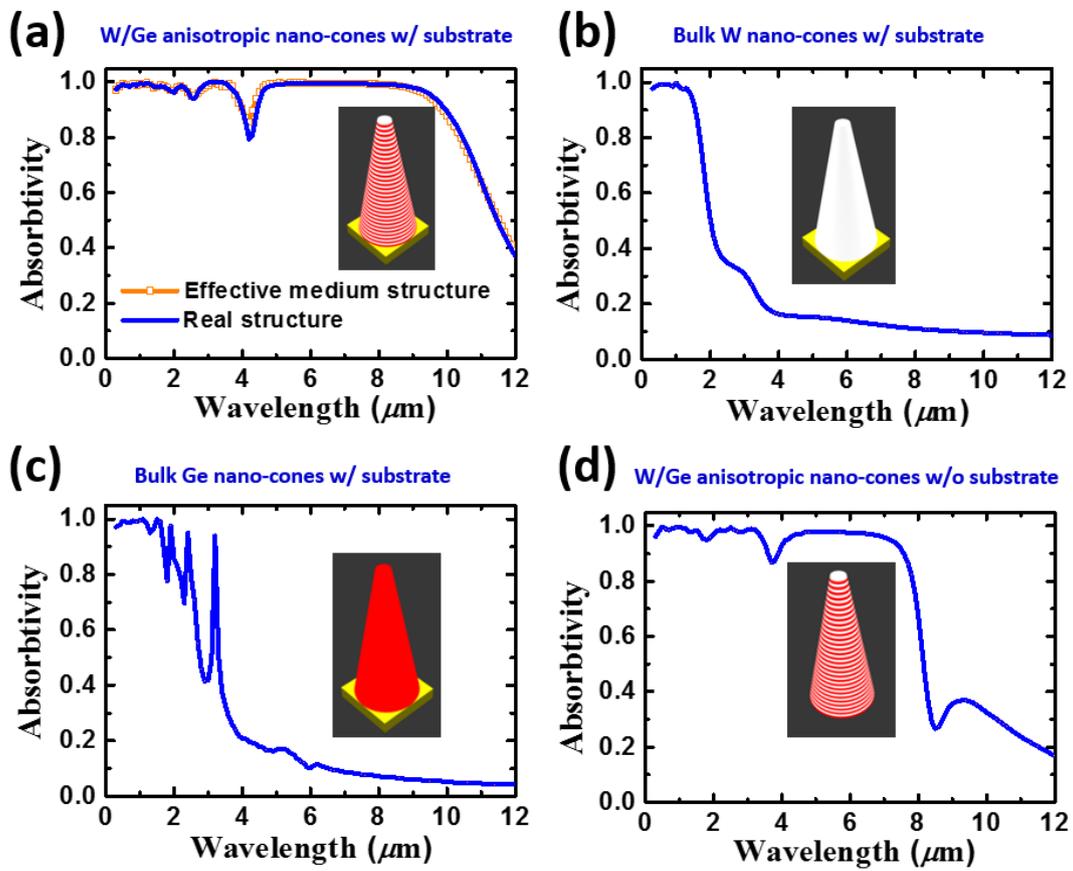

Fig. 2 Lin, Cui, et al.

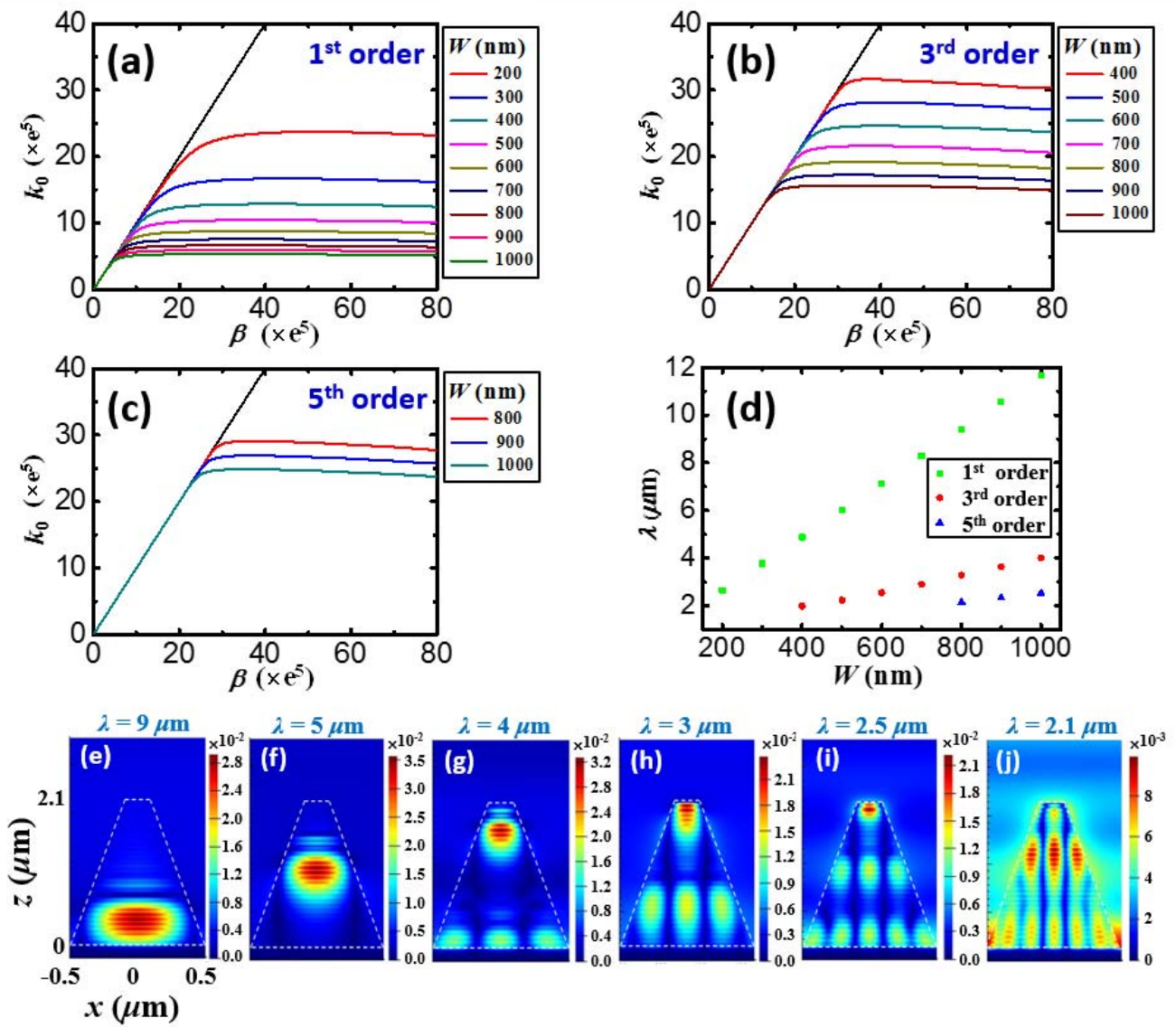

Fig. 3 Lin, Cui, et al.

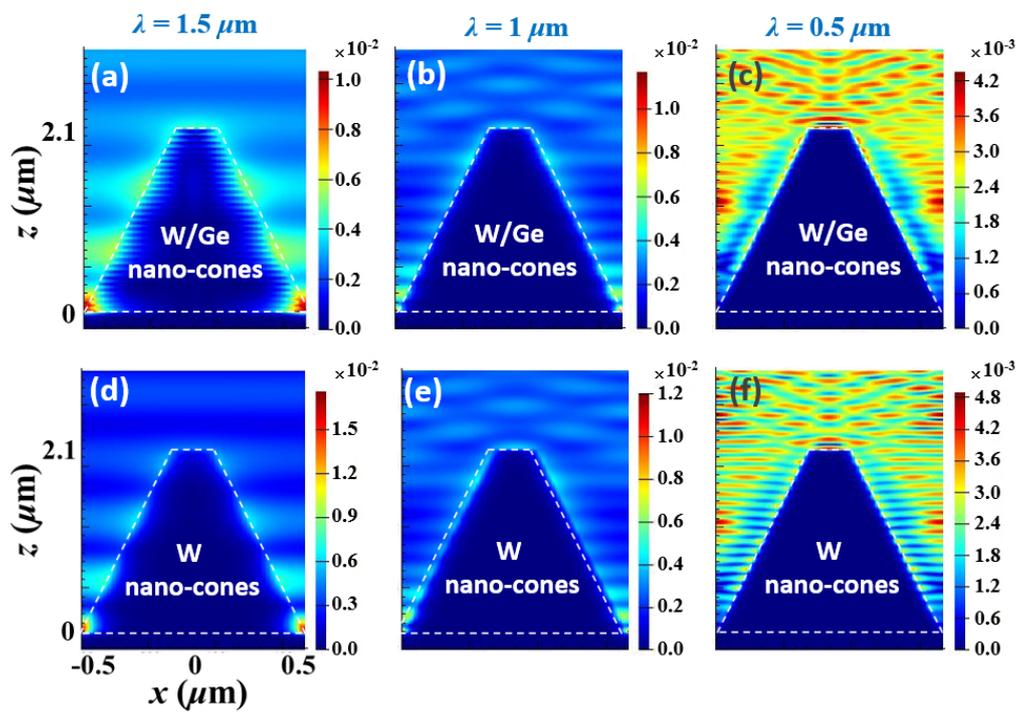

Fig. 4 Lin, Cui, et al.

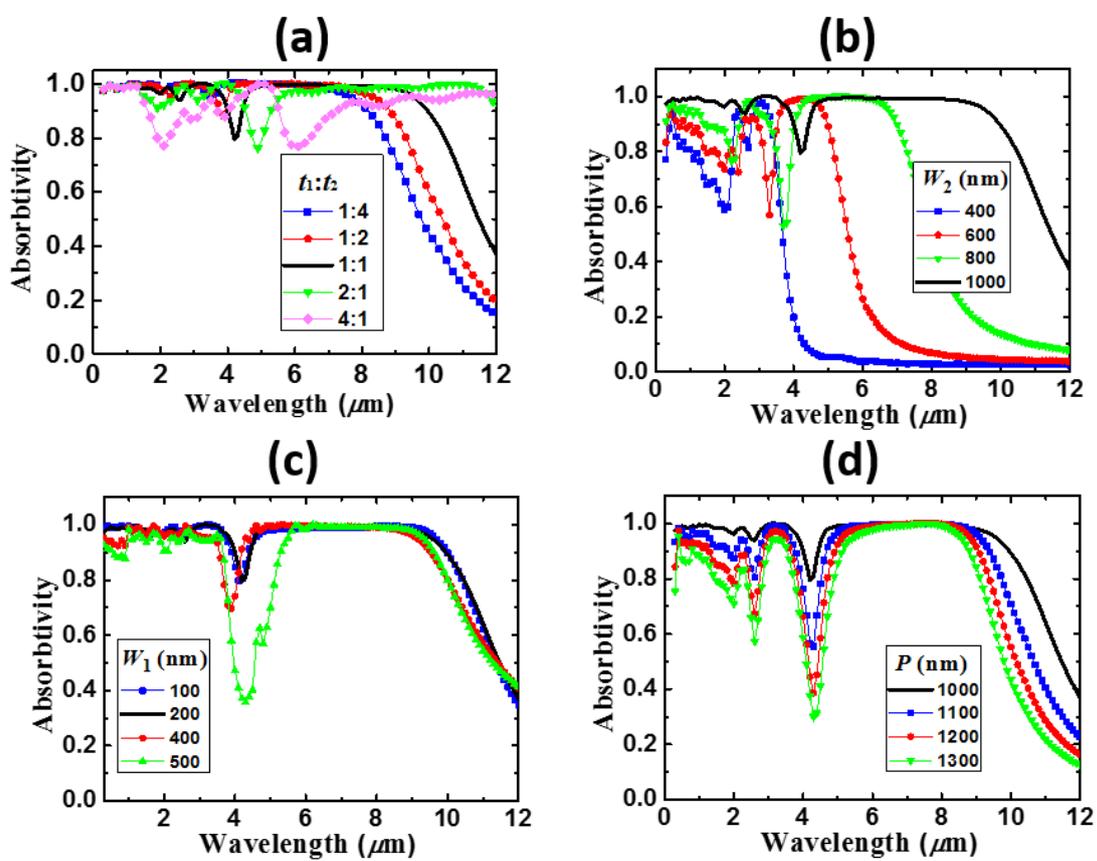

Fig. 5 Lin, Cui, et al.

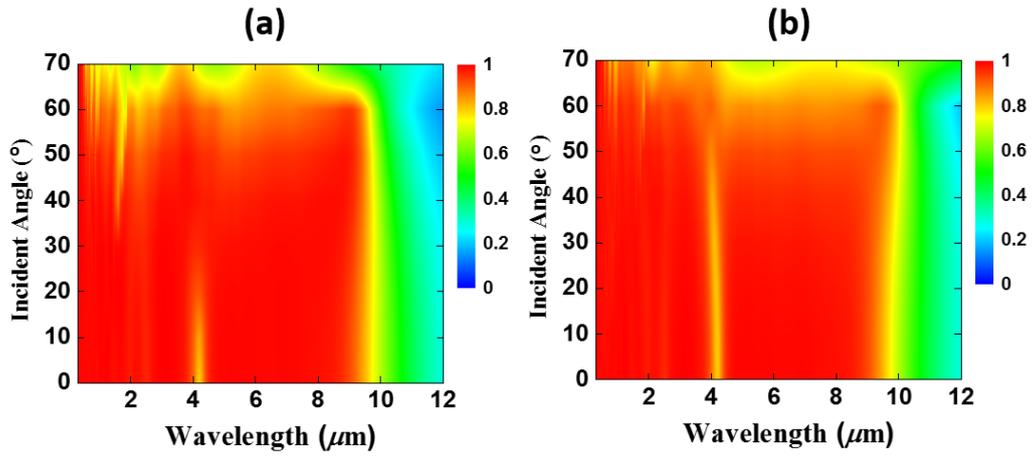

Fig. S1 Lin, Cui, et al.

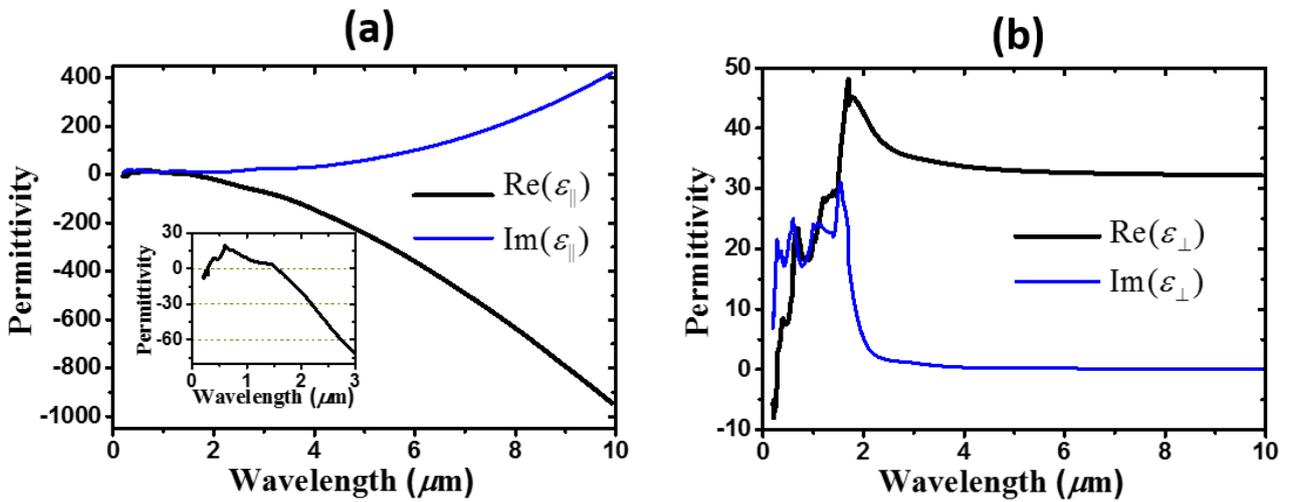

Fig. S2 Lin, Cui et al.

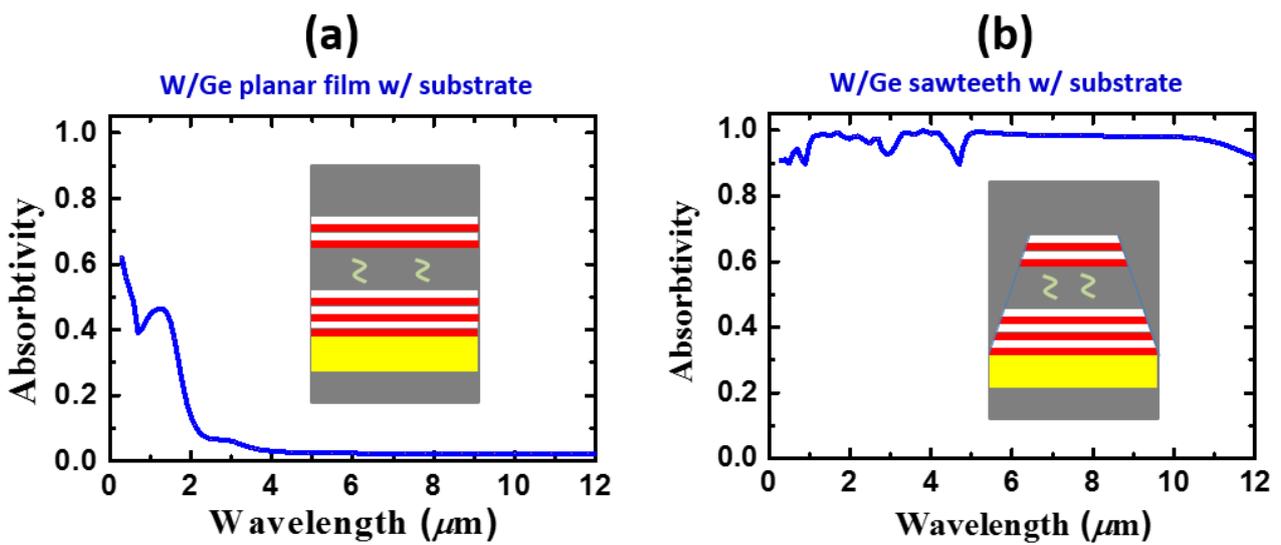

Fig. S3 Lin, Cui, et al.

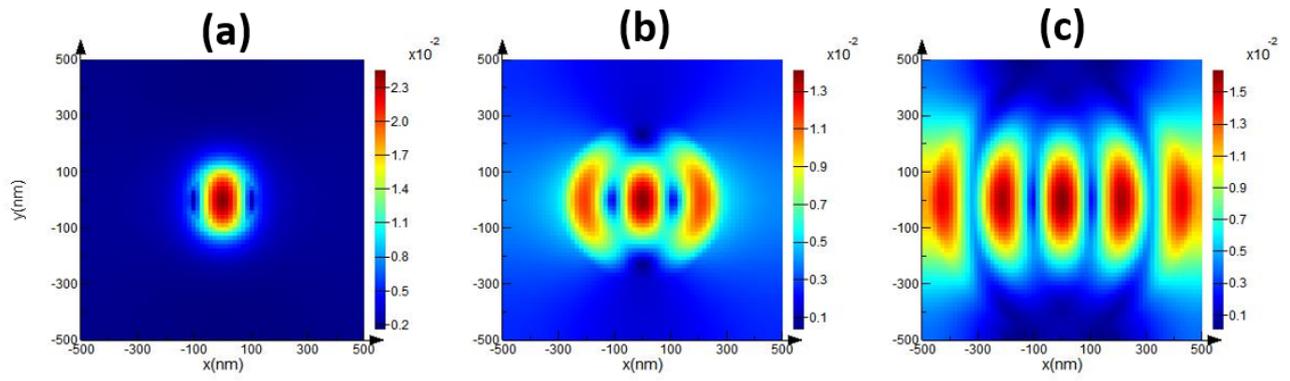

Fig. S4 Lin, Cui, et al.

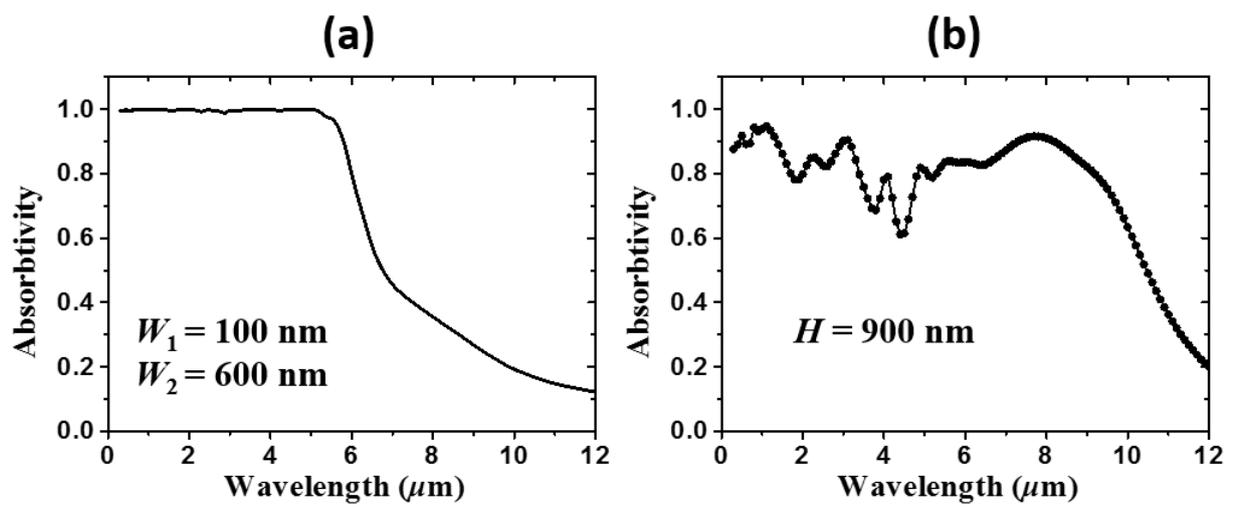

Fig. S5 Lin, Cui, et al.